\documentclass[doublecol]{epl2} %for 2 columns style without line numbers
\usepackage[dvipsnames]{xcolor}

\title{Dynamical decoupling methods in nanoscale NMR}
\shorttitle{Dynamical decoupling methods in nanoscale NMR} 

\author{C. Munuera-Javaloy \inst{1}, R. Puebla\inst{2,3} \and J. Casanova\inst{1,4}}
\shortauthor{C. Munuera-Javaloy \etal}

\institute{                    
  \inst{1} Department of Physical Chemistry, University of the Basque Country UPV/EHU, Apartado 644, 48080 Bilbao, Spain\\
\inst{2} Instituto de F{\'i}sica Fundamental, IFF-CSIC, Calle Serrano 113b, 28006 Madrid, Spain\\
\inst{3} Centre for Theoretical Atomic, Molecular, and Optical Physics, School of Mathematics and Physics, Queen’s University, Belfast BT7 1NN, United Kingdom \\
  \inst{2} IKERBASQUE, Basque Foundation for Science, Plaza Euskadi 5, 48009 Bilbao, Spain}

\abstract{
Nuclear magnetic resonance (NMR) schemes can be applied to micron-, and nanometer-sized samples by the aid of quantum sensors such as nitrogen-vacancy (NV) color centers in diamond. These minute devices allow for magnetometry of nuclear spin ensembles with high spatial and frequency resolution at ambient conditions, thus having a clear impact in different areas such as chemistry, biology, medicine, and material sciences. In practice, NV quantum sensors are driven by microwave (MW) control fields with a twofold objective: On the one hand, MW fields bridge the energy gap between NV and nearby nuclei which enables a coherent and selective coupling among them while, on the other hand, MW fields remove environmental noise on the NV leading to enhanced interrogation time. 
In this work we review distinct MW radiation patterns, or dynamical decoupling techniques, for nanoscale NMR applications. }

\begin{document}

\maketitle

\section{Introduction}
Nuclear magnetic resonance (NMR) techniques~\cite{Abragam61} are central in different scientific and technological areas due to their ability to investigate the structure of matter. In particular, there are two pivotal examples that show the wide range of application of NMR schemes. These are, on the one hand, NMR spectroscopy~\cite{Fratila11,Levitt98} which is a technique able to provide information of the different atomic element content in chemical compounds~\cite{Gunter}, as well as to determine the structure of some proteins~\cite{Fratila11, Greiner15, Duddeck98,Wuthrich90}. On the other hand, magnetic resonance imaging (MRI)~\cite{Plewes12} is a non-invasive procedure for imaging internal tissues in the human body without exposing the patient to potentially dangerous ionising radiation. 

Despite all these applications, standard NMR techniques require the inspection of large samples owing to the weak coupling between the target (i.e. a nuclear spin ensemble) and the macroscopic detectors. This results in an NMR signal that depends on the thermal  polarisation of nuclei~\cite{Bloch46} which is low at ambient conditions. 
As a figure of merit, detection of a liquid-water sample volume of $\approx$ 40 femtoliters has been achieved at room temperature~\cite{Ciobanu02}. In that experiment, a voxel of $\approx 40 \ \mu$m$^3$ (with dimensions $3.7 \pm 0.4$ $\mu$m $\times$   $3.3 \pm 0.3$ $\mu$m $\times$  $3.3 \pm 0.3$ $\mu$m) that contains $\approx 3\times 10^{12}$ protons was interrogated using micro-coil detectors, as well as strong magnetic field gradient combined with radio-frequency (rf) drivings. For more details see~\cite{Ciobanu02}. However, to further reduce conventional NMR systems to explore nanometer-sized-samples is challenging~\cite{Glover02}.

A technique that enables NMR detection at the nanoscale is magnetic resonance force microscopy (MRFM)~\cite{Sidles91} which  in 2004 achieved single electron spin detection~\cite{Rugar04}. Regarding nuclear spin detection, the reader can find a selection of MRFM experiments in Refs.~\cite{Mamin07, Mamin09, Rose18}.  In particular, in~\cite{Mamin07} a volume of the order of $\approx10^5$ nm$^3$ was scanned, which approximately contains  30 millions of nuclear spins. A similar situation can be found in~\cite{Mamin09} where it is estimated that $\sim$10$^{6}$ spins are responsible of the measured NMR signal.
In~\cite{Rose18}  a volume of $\approx (50$ nm$)^3$ of polystyrene was scanned.  Note that, for sample volumes below $(100$ nm$)^3$ the statistical nuclear polarisation would typically exceed the thermal polarisation~\cite{Degen07}, and variances in the amplitude of the exerted forces on the MRMF cantilevers are better measured. For more details regarding MRMF techniques we refer the reader to~\cite{Poggio10}. 
However, MRFM requires high vacuum conditions as well as low temperatures. This is a regime that departs from the target of achieving nanoscale NMR at ambient conditions. 

\begin{figure}
\onefigure[width=1.0\linewidth]{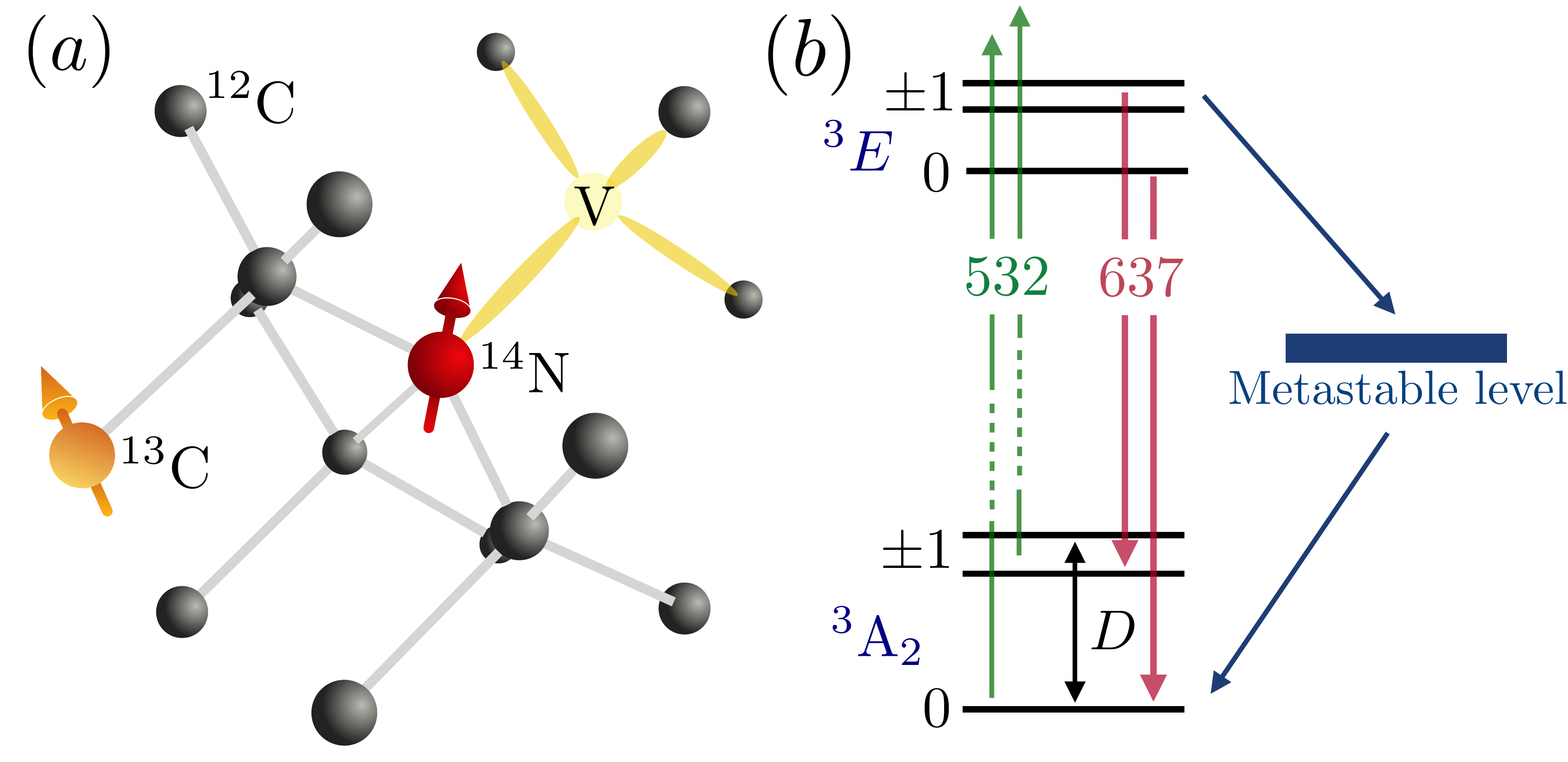}
\caption{(a) NV color center in diamond structure including a $^{14}$N atom and vacancy in the NV axis. Black atoms represent the spinless $^{12}$C nuclei, while in yellow it is represented one spin-$\frac{1}{2}$ $^{13}$C isotope  that naturally appears  with a $1.1\%$ abundance. (b) Simplified NV level structure including the electronic-ground ($^3{\rm A}_2$) and electronic-excited ($^3{\rm E}$) triplet states connected via spin conserving transitions (green and red arrows). It is also shown the presence of a transition that connects $\pm 1$ states in $^3{\rm E}$ with the 0 state in  $^3{\rm A}_2$ through a metastable level that allows to initialise the NV to 0 with high fidelity~\cite{Manson06}.}
\label{NVfig}
\end{figure} 

There are other types of imaging techniques leading to nanoscale resolution, but they also present different issues. For instance, X-ray diffraction requires the ability to crystallise a large molecular sample in an ordered manner. In addition, and similar to conventional NMR schemes, samples with  large chemical purity are needed, which poses a severe limitation. Another example is electron microscopy that also requires vacuum conditions, as well as low temperatures for the case of cryogenic electron microscopy. 

In this ecosystem of techniques, those using the {\it nitrogen vacancy (NV) center}~\cite{Doherty13} as a controllable {\it electron spin quantum sensor} stand out as they can be displayed at ambient conditions. Hence, NV-based imaging would pave the way to, e.g., investigate single molecules and proteins at physiological conditions. In this respect see, for example, Refs.~\cite{Schirhagl14, Wu16, Lovchinsky16}, as well as~\cite{Arunkumar21} and references therein. Regarding NV center structure, this is a substitutional defect in diamond where a $^{14}$N, or $^{15}$N, nitrogen isotope replaces one of the carbons in the lattice having, at the same time, an adjacent vacancy that defines the NV axis, see Fig.~\ref{NVfig} (a). Typically, the NV is identified in one of the following stable forms, namely, the negatively charged state NV$^{-}$ and the neutral state NV$^{0}$. Note that, the stable positively charged NV$^{+}$ is also being studied for different tasks~\cite{Pfender17}. In this review we focus on the NV$^{-}$ due to its suitability for nanoscale NMR applications, and hence, we will simply denote it as NV, while we refer the reader to~\cite{Doherty13} for a thorough review of the distinct NV charge states and their physical properties.

From the point of view of quantum control, NV centers can be optically polarised/initialised to the ground state of the $^3{\rm A}_2$ manifold, i.e. to the 0 level in Fig.~\ref{NVfig} (b), with a detuned laser pulse with a wavelength typically around 532 nm.  
This connects the $^3{\rm A}_2$ hyperfine triplet with that in $^3$E , green arrows in Fig.~\ref{NVfig} (b), which radiatively decays back to $^3{\rm A}_2$ emitting at $\approx 637$ nm. This optical cycle is spin conserving~\cite{Manson06}, thus no spin polarisation is gained. However, the presence of an alternative decay path through a metastable state~\cite{Manson06} that connects the $\pm 1$ hyperfine levels in $^3$E with the $0$ in $^3{\rm A}_2$ enables to initialise the NV to the 0 state of $^3{\rm A}_2$ to a degree $>92\%$ even at room temperature~\cite{Waldherr11}. It is noteworthy to mention that, at low temperatures NV polarisation $> 99\%$ has already been achieved~\cite{Robledo11}. Regarding NV readout, while the 0 and $\pm 1$ states from $^3{\rm A}_2$ radiatively decay (i.e. they emit photons) after being excited to $^3$E, $\pm 1$ states can also fall in the metastable level and remain there for $\approx 250$ ns~\cite{Manson06}. In this manner, the system gets out from the optical cycle and does not emit photons, until it decays back to the 0 level of $^3{\rm A}_2$, which enables to distinguish between 0 and $\pm 1$ states~\cite{Steiner10}. For the NV initialisation procedure described previously,  an external magnetic field $B_z$ is typically aligned with the NV axis, while the effect of magnetic field deviations with respect to the NV axis as well as the presence of nitrogen substitutional defects (i.e. $P_1$ centers) have been studied~\cite{Drake16}. 

In the absence of an external magnetic field, the $\pm 1$ states are degenerate and separated from 0 by the zero field splitting $D\approx (2\pi)\times2.87$ GHz  (cf. Fig.~\ref{NVfig} (b)). An externally applied $B_z$ field, with $z$ aligned with the NV axis, removes that degeneracy leading to the following Hamiltonian 
\begin{equation}\label{NVfree}
H = D S^2_z + |\gamma_e| B_z S_z = \omega_{1} |1 \rangle \langle 1| +  \omega_{-1} |-1 \rangle \langle -1|.
\end{equation}
Here, $S_z = |1 \rangle \langle 1| -   |-1 \rangle \langle -1|$ is a spin-1 matrix where we use the ket $|1\rangle$ ($|-1\rangle$) for the $+1$ ($-1$) state of the NV on the $^3{\rm A}_2$ hyperfine manifold. The energies $\omega_{1} = (D +  |\gamma_e| B_z)$ and $\omega_{-1} = (D -  |\gamma_e| B_z)$ with $\gamma_e =-(2\pi) \times 28.024$ GHz/T being the electronic gyromagnetic ratio, correspond to the $|1\rangle$, $|-1\rangle$ states once we set $|0\rangle$ (i.e. the ket corresponding to the 0 state of $^3{\rm A}_2$) at the zero-energy level.  Note also that in Eq.~(\ref{NVfree}) we have neglected the presence of the nitrogen spin ($^{14}$N or $^{15}$N) inherent to the NV structure (see Fig.~\ref{NVfig} (a)), as this can be polarized such that it does not participate into the NV dynamics. More specifically, the nitrogen would couple to the NV electronic spin via a term $\vec{S}\cdot {\bf A} \cdot \vec{J}$, with ${\bf A}$ being a tensor such that ${\bf A} = \rm{diag}[A^\perp, A^\perp, A^\parallel]$ and the $A^{\perp,\parallel}$ components lying on the range of few MHz (for more details see Table~1 in Ref.~\cite{Doherty13}) while $\vec{J}$ is the nitrogen spin operator. The large energy splitting of the NV, which lies in the GHz regime, effectively removes the term $A^\perp (S_x J_x + S_y J_y)$, and thus one should only take into account the dephasing-like term $A^\parallel S_z J_z$. In the case of having a $^{14}$N isotope (note this is a spin-1 nucleus with a natural  abundance larger than 99$\%$), one can polarise the nuclear spin state to $|0\rangle$ such that $J_z|0\rangle=0$, which further eliminates any contribution of the nitrogen spin to the NV dynamics~\cite{Busaite20}. In the case of deficient nitrogen polarisation or under the presence of a $^{15}$N isotope, one has to consider the term $A^\parallel S_z J_z$ as an additional source of error, although its effect may be corrected by sequentially irradiating the sample~(see below).

\section{Basic model on nanoscale NMR with NV centers} The eigenenergies $\omega_{1}$, $\omega_{-1}$ in Eq.~(\ref{NVfree}) lie in the microwave (MW) regime, thus an NV can be externally controlled with MW radiation to induce transitions between specific hyperfine levels.  These tools, namely efficient spin initialisation, readout, and control via optical and MW radiation, enable the detection and control of nearby nuclear spins even at room temperature as it was firstly demonstrated by Jelezko et al.~\cite{Jelezko04, Jelezko04bis}, and Childress et al.~\cite{Childress06}. Note that, recently, electrical control of an NV center has also been demonstrated~\cite{Gulka21}.

The scenario involving an NV coupled to a nuclear spin bath and driven by MW radiation is described by the following Hamiltonian ($\hbar=1$)
\begin{eqnarray}\label{basic}
H &=& \omega_{1} |1 \rangle \langle 1| +  \omega_{-1} |-1 \rangle \langle -1| - \sum_j \gamma_n B_z I_j^z\nonumber\\
&+& H_{\rm NVn} + H_{\rm nn} + H_{\rm driv}.
\end{eqnarray}
In the previous expression, the term  $- \sum_j \gamma_n B_z I_j^z$ accounts for the Zeeman splitting of each nuclei as a consequence of the externally applied $B_z$ field. In addition, $I_j^z$ is the nuclear spin operator and  $\gamma_n$ the nuclear gyromagnetic ratio. Note that, for the sake of simplicity in the presentation, we consider an homonuclear sample. The term  $H_{\rm NVn}$ describes the interaction between the NV and each nucleus~\cite{Maze08} and it will deemed as dipole-dipole leading to   
\begin{equation}
H_{\rm NVn} = \sum_j \frac{\mu_0 \gamma_e \gamma_n}{4 \pi |\vec{r}_j|^3} \bigg[\vec{S}\cdot \vec{I}_j - \frac{3 \big(\vec{S}\cdot \vec{r}_j\big)\big(\vec{I}_j\cdot \vec{r}_j\big)}{|\vec{r}_j|^2}\bigg],
\end{equation}
where the vector $\vec{r}_j$ determines the position of each nucleus with respect to the NV. In the same manner, $H_{\rm nn}$ accounts for dipolar interactions among nuclei~\cite{Casanova16} ($\vec{r}_{j,k}$ is the vector that connects the $j$th and $k$th nuclei), and can be written as
\begin{equation}
H_{\rm nn} = \sum_{j<k} \frac{\mu_0 \gamma^2_n}{4 \pi |\vec{r}_{j,k}|^3} \bigg[\vec{I}_j\cdot \vec{I}_k - \frac{3 \big(\vec{I}_j\cdot \vec{r}_{j,k}\big)\big(\vec{I}_k\cdot \vec{r}_{j,k}\big)}{|\vec{r}_{j,k}|^2}\bigg].
\end{equation}
Finally, the MW driving acting on the NV reads as
\begin{equation}\label{driv}
  H_{\rm driv}=\sqrt{2}\Omega S_x \cos{(\omega t -\phi)},
\end{equation}
where $\Omega$ is the Rabi frequency, $\phi$ the driving phase, and the factor $\sqrt{2}$ has been introduced by convenience. In later developments, we will refer to $\phi$ as the pulse phase.

\section{Continuous schemes and the Hartmann-Hahn double resonance condition} In the absence of a MW driving, the Hamiltonian in Eq.~(\ref{basic}) does not enable the exchange of interactions between NV and nuclei as a consequence of the large energy difference among them. Note that the NV involves transitions on the range of GHz, see definitions of $\omega_{\pm 1}$ in the previous section, while the nuclear splitting is, typically, in the range of hundreds of kHz to some MHz depending on the value of $B_z$. 

More specifically, in the rotating frame of $ \omega_{1} |1 \rangle \langle 1| +  \omega_{-1} |-1 \rangle \langle -1| - \sum_j \gamma_n B_z I_j^z$ (i.e. the first line in Eq.~(\ref{basic})) and removing fast rotating terms by invoking the rotating wave approximation (RWA), the $H_{\rm NVn}$ term approximately simplifies to  
\begin{equation}\label{ZZ}
H_{\rm NVn} \approx \sum_j A_j^z S_z I_j^z, 
\end{equation} 
with $A_j^z$ being the $z$ component of the hyperfine vector $\vec{A}_j = (A_j^x,  A_j^y, A_j^z) =\frac{\mu_0\gamma_e\gamma_n}{4\pi |\vec{r}_j|^3}[\hat{z} - 3\frac{(\hat{z}\cdot\vec{r}_j)\vec{r}_j}{|\vec{r}_j|^2}]$. It is worth noting that Eq.~(\ref{ZZ}) describes the loss of coherence of the the NV degrees of freedom due to the uncontrolled  interaction with the nuclear environment. In order to achieve coherent interactions between the NV and some specific nuclei one can introduce a {\it continuous MW driving}  
resonant, e.g., with the $|0\rangle \leftrightarrow |1\rangle$ NV electron spin transition. Then, in a rotating frame with respect to (w.r.t.) 
$\omega_{1} |1 \rangle \langle 1| +  \omega_{-1} |-1 \rangle \langle -1|$, and noting that $S_z = \frac{1}{2} (\sigma_z  + I) - |-1\rangle\langle -1|$ where
$\sigma_z = |1\rangle\langle 1| - |0\rangle\langle 0|$ and $I=|1\rangle\langle 1| + |0\rangle\langle 0|$ Eq.~(\ref{basic}) reads
\begin{eqnarray}\label{updatedmodel}
H &=& - \sum_j \omega_j \hat{\omega}_j \cdot \vec{I}_j + \frac{\sigma_z}{2} \sum_j \vec{A}_j \cdot \vec{I}_j + H_{\rm nn}\\ \nonumber
&&+ \frac{\Omega}{2} \big(|1\rangle\langle0| e^{i\phi} + |0\rangle\langle 1| e^{-i\phi}\big).
\end{eqnarray} 
Here, the nuclear resonance is given by $\omega_j = |\vec{\omega}_j| \approx |\gamma_{n} B_z -\frac{1}{2} A_j^z|$, where $\vec{\omega}_j =  \gamma_{n} B_z\hat{z} - \frac{1}{2} \vec{A}_j$ and $\hat{\omega}_j = \vec{\omega}_j/\omega_j$. The MW driving is set on resonance with the energy of the $|0\rangle \leftrightarrow |1\rangle$ transition, thus the $|-1\rangle$ NV component is decoupled I think decoupled is more appropriate than removed from the dynamics, while $H_{\rm nn}$ can be simplified to a very good approximation to 
\begin{equation}
H_{\rm nn} \approx \sum_{j<k} \frac{\mu_0 \gamma^2_n}{4 \pi |\vec{r}_{j,k}|^3} (1-3 (n^z_{j,k})^2)\bigg[ I^z_jI^z_k - (I^+_jI^-_k + I^-_jI^+_k )\bigg]
\end{equation}
once that fast rotating terms are removed by the RWA. In this expression, we have defined $I^{\pm} = \frac{1}{2}\left(I^x\pm i I^y\right)$. Now, with the help of the following identity,
\begin{equation}\label{identity}
e^{i\vec{I}_{j}\cdot\hat{l}\phi}\vec{I}_{j}\cdot\vec{b}\ e^{-i\vec{I}_{j}\cdot\hat{l}\phi}=\vec{I}_{j}\cdot[(\vec{b}-\vec{b}\cdot\hat{l}\hat{l})\cos\phi-\hat{l}\times\vec{b}\sin\phi+\vec{b}\cdot\hat{l}\hat{l}],
\end{equation}  
one can write Eq.~(\ref{updatedmodel}) in the rotating frame of $- \sum_j \omega_j \hat{\omega}_j \cdot \vec{I}_j  + \frac{\Omega}{2} \big(|1\rangle\langle0| e^{i\phi} + |0\rangle\langle 1| e^{-i\phi}\big)$ as
\begin{eqnarray}
H &=& \frac{1}{2}\big( |+_\phi \rangle\langle -_\phi| e^{i\Omega t} + {\rm H.c.} \big)\sum_j \vec{I}_j\cdot \big[A^x_j \hat{x}_j \cos{(\omega_j t)}\nonumber\\
&-& A^y_j \hat{y}_j \sin{(\omega_j t)} +  A^z_j \hat{z}_j\big] + \tilde{H}_{\rm nn}.
\end{eqnarray}
Here $|\pm_\phi \rangle = \frac{1}{\sqrt{2}}(|1\rangle \pm e^{-i\phi} |0\rangle)$, $A^x_j = | \vec{A}_j - (\vec{A}_j\cdot \hat{\omega}_j) \hat{\omega}_j |$, $\hat{x}_j = \big[\vec{A}_j - (\vec{A}_j\cdot \hat{\omega}_j) \hat{\omega}_j \big]/ A^x_j$, $A^y_j =| \hat{\omega}_j \times \vec{A}_j |$,  $\hat{y}_j =  - (\hat{\omega}_j \times \vec{A}_j)/A^y_j$, and $A^z_j = |(\vec{A}_j\cdot \hat{\omega}_j) \hat{\omega}_j|$, $\hat{z}_j = (\vec{A}_j\cdot \hat{\omega}_j) \hat{\omega}_j/A^z_j$. Note also that one can easily demonstrate that $A^x_j=A^y_j = A^\perp_j$, while we refer to $A^\perp_j$ as the {\it perpendicular $j$th coupling}.

In particular, when $\Omega = \omega_l$, where $l$ labels one of the nuclear resonances of the system (i.e. $\omega_l \in \{\omega_j\}$)  we achieve the so-called {\it Hartmann-Hahn (HH) resonance condition~\cite{Hartmann62}} that enables a flip-flop interaction, i.e. exchange excitations, between the NV and the $l$th nucleus, 
\begin{equation}\label{HHHamiltonian}
H \approx \frac{A^{\perp}_l}{2} \big[|+_\phi \rangle\langle -_\phi| I^{-}_l + |-_\phi \rangle\langle +_\phi| I^{+}_l  \big].
\end{equation}

The continuous driving $H_{\rm driv}=\sqrt{2}\Omega S_x \cos{(\omega t -\phi)}$ (cf. Eq.~(\ref{driv})), that was introduced in Eq.~(\ref{basic}) has a twofold intention. On the one hand, $H_{\rm driv}$ allows for a coherent coupling between the NV and specific nuclei and, on the other hand, it can be employed to  remove the effect of environmental noise over the quantum sensor. In the case of Eq.~(\ref{HHHamiltonian}) this would enter as a coupling with detuned nuclear spins in the environment.  In this manner, the $H_{\rm driv}$ term is our first example of dynamical decoupling (DD) techniques~\cite{Souza12}. 

The flip-flop term that has been derived in Eq.~(\ref{HHHamiltonian}) allows for a sequential transfer of excitations from an optically initialized NV, or from an NV ensemble, to the surrounding nuclear spin bath~\cite{Chen15}. It is important to note that, after NV depolarisation due to the exchange mechanism in Eq.~(\ref{HHHamiltonian}), the NV can be optically repolarised, enabling thus a continuous transfer of excitations to the nuclear bath.  Such procedure would result in an enhancement of the sample polarisation, and consequently of the NMR signal, leading to larger NMR sensitivities. The procedure of exchanging excitations among NVs and nuclei can be cast in the general frame of {\it dynamic nuclear polarisation} (DNP) schemes. The reader can find introductory references to DNP techniques in~\cite{Abragam78, Prisner08, Atsarkin11, Corzilius20}.  In particular, the scheme  in Eq.~(\ref{HHHamiltonian}) was demonstrated in NV-based systems by London et al. in Ref.~\cite{London13}. This experimental demonstration led to polarisation transfer between the NV and $^{13}$C nuclei mediated by a continuous driving that holds the HH condition. Further developments using MW-mediated  DNP techniques  among optically-polarised-NVs and nuclei have also been reported~\cite{Alvarez15, King15}. In addition, while polarisation transfer protocols typically require a fine alignment of $B_z$ with the NV axis, a scheme combining laser and MW drivings to mitigate the impact of a large misalignment has been put forward in Ref.~\cite{Ajoy18}. This method resulted in the design of a miniaturised {\it hyperpolariser} that has delivered diamond particles with a $0.86\%$ of $^{13}$C~\cite{Ajoy20}.

Previously, other NV-based polarising schemes were performed. For example, in Ref.~\cite{Gurudev07} a single $^{13}$C nucleus was polarised with selective MW drivings combined with the tracking of the nuclear spin precession, while in Ref.~\cite{Jacques09} the use of excited-state-level anticrossing (ESLAC) led to nuclear spin polarisation without MW fields. Subsequently, a nuclear polarisation mechanism based on ground-state-level anticrossing (GSLAC) was demonstrated~\cite{Wang13}. In addition, note that several techniques based on the physics of  level anticrossing (LAC) have been demonstrated leading to bulk $^{13}$C polarisation~\cite{Fischer13, Pagliero18}.  

Beyond standard continuous MW irradiation patterns of the kind $H_{\rm driv}=\sqrt{2}\Omega S_x \cos{(\omega t -\phi)}$, other  and more sophisticated continuous MW schemes have been put forward for different tasks. In particular, concatenated continuous dynamical decoupling (CCD) techniques use several MW tones to obtain enhanced robustness since environmental errors and deviations on the controls can be suppressed up to first order~\cite{Cai12}. We introduce the CCD technique via the following Hamiltonian that describes the interaction of an NV with two MW drivings, 
\begin{eqnarray}\label{CCD}
H &=& \frac{\omega_0 + \Delta}{2}\sigma_z + \Omega_1 (1 +\xi_1) \sigma_x \cos(\omega_1 t) \nonumber\\
&+&  \Omega_2 (1 +\xi_2) \sigma_x \cos(\omega_1 t -\phi)\cos(\Omega_1 t).
\end{eqnarray}
Here, $\omega_0$ is the two-level energy gap (in the case of an NV $\omega_0 = D + |\gamma_e| B_z$ when considering the qubit defined by the $|0\rangle \leftrightarrow |1\rangle$ transition), $\Omega_{1,2}$ are the Rabi frequencies of the MW drivings such that $|\Omega_1|\gg|\Omega_2|$, while $\Delta\equiv\Delta(t)$ and $\xi_{1,2}\equiv\xi_{1,2}(t)$ define environmental and control errors that we aim to cancel or minimize. Note that the error on Rabi frequencies is proportional to the MW field amplitude, that is, $\Omega_{1,2}\xi_{1,2}$. The simplest CCD method employs two MW driving fields: The first one with frequencies $\omega_1$ and $\Omega_1$ (second term at the right-hand-side of the first line in Eq.~(\ref{CCD})) and a second MW driving with frequencies $\omega_1$, $\Omega_1$ (second line in Eq.~(\ref{CCD})). These MW driving fields mitigate the aforementioned errors in the following manner: In a rotating frame w.r.t $\frac{\omega_0}{2}\sigma_z$ and setting $\omega_1=\omega_0$, Eq.~(\ref{CCD}) transforms to  
\begin{eqnarray}\label{CCD2}
H = \frac{\Delta}{2}\sigma_z + \frac{\Omega_1}{2} \sigma_x  + \frac{\Omega_1\xi_1}{2} \sigma_x + \frac{\Omega_2}{2} (1 +\xi_2) \sigma_\phi\cos(\Omega_1 t)
\end{eqnarray}
where $\sigma_\phi = \sigma^+ e^{i\phi} + \sigma^- e^{-i\phi}$. In this frame, the driving $\frac{\Omega_1}{2} \sigma_x$ can be understood as the free energy term of the qubit which is now defined in the dressed state basis $|\pm_x\rangle= \frac{1}{\sqrt{2}}(|1\rangle \pm |0\rangle)$. Interestingly, the error term $\frac{\Delta}{2}\sigma_z$ that was parallel to the original qubit basis in Eq.~(\ref{CCD}), is now perpendicular to  $\frac{\Omega_1}{2} \sigma_x$ and thus, it induces transitions in this new basis. However, if $\Omega_1 \gg \Delta$ these unwanted transitions suffer an energy penalty and can be effectively removed. Furthermore, in a rotating frame w.r.t. $\frac{\Omega_1}{2} \sigma_x$ and setting $\phi=-\pi/2$, Eq.~(\ref{CCD2}) becomes
\begin{eqnarray}\label{CCD3}
H \approx \frac{\Omega_1 \xi_1}{2}  \sigma_x + \frac{\Omega_2}{4} \sigma_y  +\frac{\Omega_2\xi_2 }{4} \sigma_y.
\end{eqnarray}
Similar to the previous step, now we have $\frac{\Omega_2}{4} \sigma_y$ that redefines the qubit-free-energy term in the basis  $|\pm_y\rangle$  ($\sigma_y |\pm_y\rangle = \pm  |\pm_y\rangle$). In this manner, the effect of the noisy term $\frac{\Omega_1 \xi_1}{2}  \sigma_x$ can be neglected if $|\Omega_2| \gg |\Omega_1 \xi_1|$. Hence, the final Hamiltonian adopts a simple form, 
\begin{eqnarray}\label{CCD4}
H \approx \frac{\Omega_2}{4} \sigma_y  +\frac{\Omega_2\xi_2 }{4} \sigma_y.
\end{eqnarray}
This CCD technique allows for the simultaneous mitigation of environmental errors, denoted by $\Delta$, as well as deviations from the first driving, i.e. $\Omega_1 \xi_1$. Note that the remaining error contribution, given by $\Omega_2\xi_2$, is small as we assumed $|\Omega_1|\gg|\Omega_2|$. The CCD scheme has already been demonstrated in NV centers leading to improved coherence times in the system~\cite{Cai12}.

Continuous DD techniques including MW fields with several tones have been also proposed as a mechanism to couple NVs to fast oscillating targets~\cite{Casanova19}.  In particular, if one introduces in Eq.~(\ref{basic}) a MW driving of the kind 
\begin{equation}\label{bridgeHH}
H_{\rm driv}=\sqrt{2}\Omega(t) S_x \cos{(\omega t)}
\end{equation}  
such that $\Omega(t) = \Omega_0 - \Omega_1 \sin(\nu t)$, one can easily derive the following Hamiltonian that governs the NV-nuclei interaction, 
\begin{eqnarray}\label{JA}
H &=& \frac{1}{2} [|+\rangle\langle -| e^{i\Omega_0 t} e^{i(\Omega_1/\nu) \cos{(\nu t)}}  + {\rm H.c.}]     \nonumber\\
&&\times\sum_j \vec{I}_j\cdot \big[A^x_j \hat{x}_j \cos{(\omega_j t)} - A^y_j \hat{y}_j \sin{(\omega_j t)}  \nonumber\\
&&+  A^z_j \hat{z}_j\big] + \tilde{H}_{\rm nn}.
\end{eqnarray}
Now, by using the Jacoby-Anger expansion --i.e. $e^{iz\cos(\theta)} = J_0(z)  + 2\sum_{n=1}^{+\infty}i^n J_n(z)\cos{(n\theta)}$, where $J_n(z)$ is the $n$th-order Bessel function of the first kind-- one can find that for $\Omega_0 + \nu = \omega_j$ Eq.~(\ref{JA}) reads as
\begin{equation}\label{lowcoupling}
H\approx (A^\perp_j/2)J_1(\Omega_1/\nu) ( i|+\rangle\langle -| I^+_j - i|-\rangle\langle +| I^-_j).
\end{equation}
The previous Hamiltonian defines a coherent interaction between an NV and a specific nucleus beyond the HH condition.  Note that, in the case of having a MW driving with a constant Rabi frequency, as considered in Eq.~(\ref{driv}), the HH condition imposes  $\Omega = \omega_j$. However, a suitable time modulation of $\Omega(t)$ leads to the alternative condition $\Omega_0 + \nu = \omega_j$. Such modified condition allows for the communication between NVs and nuclei through Eq.~(\ref{lowcoupling}) with low power MW radiation patterns~\cite{Casanova19}, since a small value for $\Omega_0$ (which implies low power in the driving as this is $\propto |\Omega_0|^2$) is compensated by the frequency $\nu$. The latter is responsible of modulating the Rabi frequency without significantly changing the energy amount. 
\section{Pulsed schemes} MW radiation can also be stroboscopically delivered to the system in the form of pulse sequences. Typically, these encompass trains of $\pi$-pulses that flip the NV electron spin according to $\sigma_z \rightarrow -\sigma_z \rightarrow \sigma_z \rightarrow -\sigma_z \cdots$, while $\pi/2$ pulses can be  interspersed (see below).  
In this scenario, it is common to describe the basic model in Eq.~(\ref{updatedmodel}) within the rotating frame of $\frac{\Omega}{2} \big(|1\rangle\langle0| e^{i\phi} + |0\rangle\langle 1| e^{-i\phi}\big)$ which leads to 
 \begin{equation}\label{modulated}
H = - \sum_j \omega_j \hat{\omega}_j \cdot \vec{I}_j + F(t) \frac{\sigma_z}{2} \sum_j \vec{A}_j \cdot \vec{I}_j + H_{\rm nn}.
\end{equation} 
Here, $F(t) = \pm1$ is a modulation function that describes the action of each $\pi$ pulse on the dynamics. Note that, for the sake of simplicity in the presentation, we consider {\it instantaneous pulses}, while we refer the interested reader to Ref.~\cite{Lang17} for a comprehensive study of the effect of  non-negligible pulse-width as well as of their associated applications in nanoscale NMR. Now, the role of the {\it pulse-phase} $\phi$ becomes clear since it determines the {\it rotation axis} along which the $\pi$ pulse is performed. Paradigmatic examples are: (i) The case of $\phi=0$ where the spin flip is performed along the x axis leading to an X pulse, and (ii) $\phi=-\pi/2$ leading to a Y pulse (i.e. the rotation occurs along the y axis). Later we will comment on the need of appropriately selecting pulse phases as these can result in improved sequence robustness. 

\begin{figure}
\onefigure[width=1.0\linewidth]{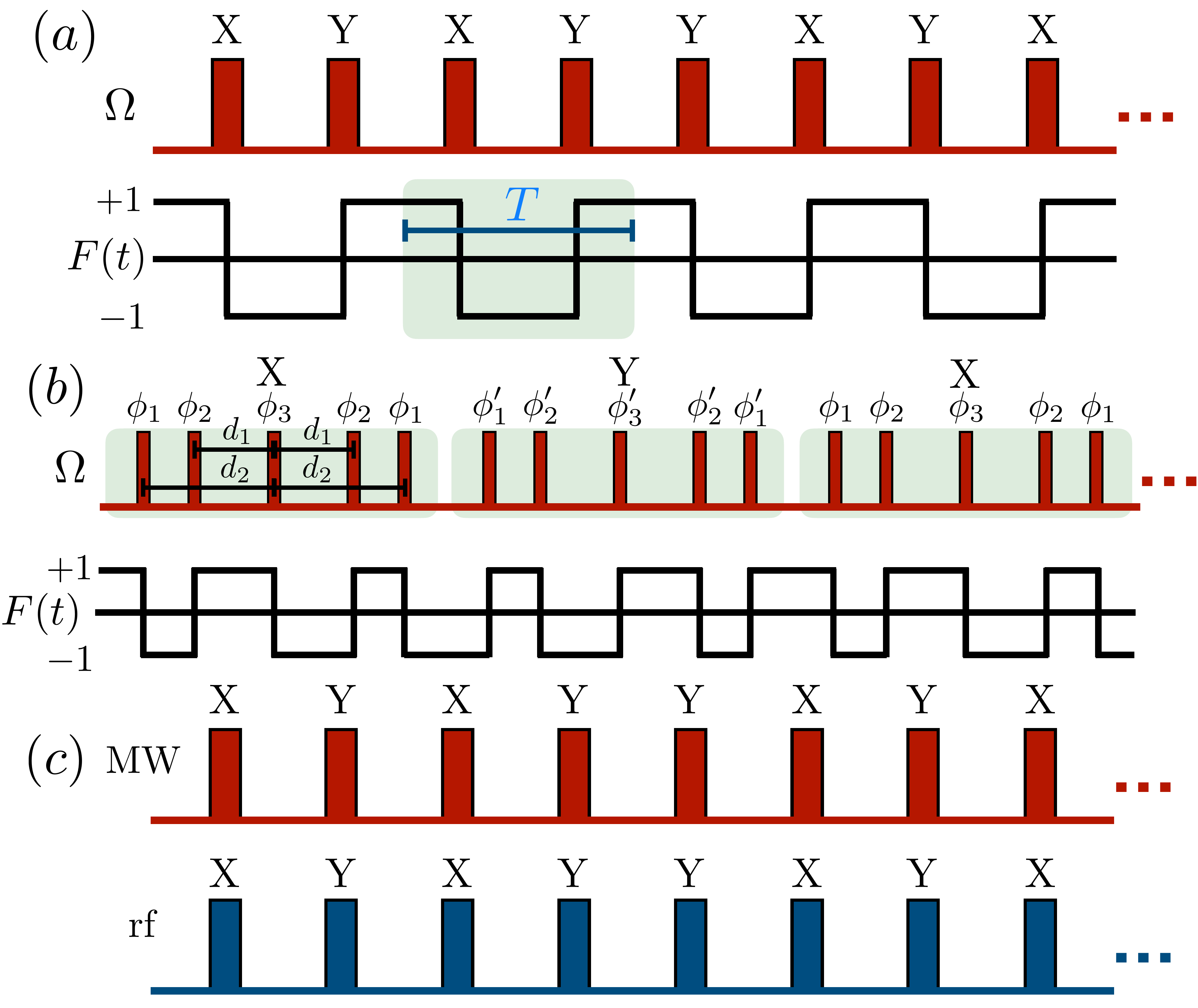}
\caption{(a) First panel shows a XY8 pulse block. Note that an XY8 sequence will be [XYXYYXYX$]^N$ with N being the number of repetitions. Second panel, modulation function $F(t)$ induced by the XY8 sequence and its associated period $T$. (b) First panel illustrates the first composite pulses of an AXY8 sequence showing the tunability of the interpulse distances. Note the internal symmetry of the composite pulse where the distance between the first $\pi$ pulse and the central one equals that of the central $\pi$ pulse with the last one. That symmetry is also present among the second-central and third $\pi$ pulses and leads to extra robustness against control errors, see Ref.~\cite{Casanova15} for further details. The pulse phase on each block is $\phi_1=\pi/6$, $\phi_2=0$, $\phi_3=\pi/2$, while $\phi_1'=\phi_1 +\pi/2$, $\phi_2'=\phi_2 +\pi/2$, and $\phi_3'=\phi_3 +\pi/2$ for enhanced robustness~\cite{Casanova15}. Second panel, modulation function resulting from the application of the sequence in the first panel. (c) Synchronised delivery of MW and rf pulses for exploiting the parallel NV-nucleus coupling. After MW and rf pulses the parallel interaction does not change leading to the same $e^{-i t A^\parallel_j \sigma_z I_u^z}$ propagator. This owes to the combined sign change on the $\sigma_z$ and $I^z_u$ operators after synchronised pulses. Other nuclei, not resonant with the rf pulse, do not flip which ultimately results in a cancellation of their dynamics.}
\label{pulsesfig}
\end{figure}

On the one hand, the delivery of $\pi$ pulses, leading to $F(t)$ in Eq.~(\ref{modulated}), may be capable of extending the NV coherence by removing the interactions with nearby nuclear spins.  This has been demonstrated using different pulse sequences both at room-,  and low-temperatures~\cite{Ryan10, Lange10, BarGill13, Abobeih18}. In these previously mentioned experiments 
different types of sequences, including  pulses with distinct pulse phases, have been tested. Among them, one finds the Carr-Purcell-Meiboom-Gill (CPMG) sequence that includes blocks of X pulses~\cite{Carr54, Meiboom58}, the Uhrig DD sequence~\cite{Uhrig07} that extends the CPMG scheme, and the XY8 sequence, 
that belongs to the XY family of pulse schemes~\cite{Maudsley86, Gullion90} and consists on sequentially repeating the block XYXYYXYX. As an example, an XY8 block is depicted in Fig.~\ref{pulsesfig} (a) first panel.

On the other hand, the appropriate tuning of $F(t)$ enables us to detect and control nuclear spins. This can be  seen when Eq.~(\ref{modulated}) is expressed in a rotating frame with respect to $- \sum_j \omega_j \hat{\omega}_j \cdot \vec{I}_j$, which adopts the following form
\begin{eqnarray}\label{pulsedH}
H &=& F(t) \frac{\sigma_z}{2} \sum_j \vec{I}_j\cdot \big[A^x_j \hat{x}_j \cos{(\omega_j t)}\nonumber\\
&-& A^y_j \hat{y}_j \sin{(\omega_j t)} +  A^z_j \hat{z}_j\big] + \tilde{H}_{\rm nn}.
\end{eqnarray}
In Fig.~\ref{pulsesfig} (a) second-panel, the modulation function imprinted by an XY8 sequence is shown. This is a periodic function with period $T$ and presents an {\it even} form such that it can be expanded in cosine-like {\it harmonics} as $F(t) = \sum_k f_k \cos{(k \frac{2\pi}{T} t)}$ with $f_k$ being the $k$th Fourier coefficient 
$f_k = \frac{2}{T} \int_0^T F(t) \cos{(k \frac{2\pi}{T} t)} dt$. In this manner, by selecting the interpulse spacing (thus the period $T$) one can induce a resonance in Eq.~(\ref{pulsedH}) leading to a coherent NV-nucleus dynamics. In particular, when $q \frac{2\pi}{T}  = \omega_u$ (i.e., if one sets the period $T$ such that  the $q$th harmonic is resonant with the $u$th nucleus) Eq.~(\ref{pulsedH}) simplifies to 
\begin{equation}\label{effectiveNVN}
H \approx \frac{f_q A_u^\perp}{4} \sigma_z I_u^x + \tilde{H}_{\rm nn},
\end{equation}
enabling the detection and control of the $u$th nucleus via the NV center. In the same manner, and by selecting an {\it odd} distribution of $\pi$ pulses (at this respect see, e.g., Ref.~\cite{Casanova16}) such that the resulting $F(t) = \sum_k \tilde{f}_k \sin{(k \frac{2\pi}{T} t)}$ with $\tilde{f}_k = \frac{2}{T} \int_0^T F(t) \sin{(k \frac{2\pi}{T} t)} dt$ one would obtain the following interaction $H \approx \frac{\tilde{f}_q A_u^\perp}{4} \sigma_z I_u^y + \tilde{H}_{\rm nn}$. Interestingly, the type of sequences involving a constant interpulse spacing (e.g., the one in Fig.~\ref{pulsesfig} (a) first-panel) lead to a $F(t)$ such that $f_q = \frac{4}{q\pi}$. That is, the effective NV-nucleus strength, i.e. $\frac{f_q A_u^\perp}{4}$ in Eq.~(\ref{effectiveNVN}),  can be modulated by selecting different harmonics as this changes the value of $q$. This approach was used in several experiments during the last decade to detect nuclear spins that are weakly coupled with the NV, that is,  nuclear spins with a coupling strength smaller than the NV {\it bare dephasing rate} (i.e. the NV dephasing rate without being driven by DD techniques). The reader can find several experiments exploiting the previously mentioned pulsed DD techniques for {\it individual nuclear spin control} in Refs.~\cite{Taminiau12, Kolkowitz12, Zhao12, Muller14, Lang19, Bradley19}, as well as for the detection of {\it nuclear spin clusters} in Refs.~\cite{Zhao11, Shi14, Abobeih19}. Furthermore, NV-based magnetic field detection can be combined with an external clock leading to a large sensitivity as it was demonstrated in Refs.~\cite{Schmitt17, Glenn18, Chu21, Mizuno18}.

Other DD pulse sequences that involve tunable interpulse spacings have been proposed. For instance, the adaptive-XY8 (AXY8) sequence~\cite{Casanova15} contains eight composite X and Y pulses ordered as XYXYYXYX, where each X and Y includes five $\pi$ pulses with adjustable distances. In Fig.~\ref{pulsesfig} (b), first-panel, the first three composite pulses of an AXY8 sequence are shown. There it can be seen the two interpulse distances $d_1$ and $d_2$ that can be arbitrarily selected leading to different modulation functions such as the one depicted in Fig.~\ref{pulsesfig} (b) second-panel. The tunability of $F(t)$ provides a large nuclear spin selectivity~\cite{Casanova15}, while the case $d_1=d_2$ (i.e., equidistant $\pi$ pulses) coincides with the KDD$_\phi$-KDD$_{\phi + \pi/2}$ sequence~\cite{Souza11}. It is noteworthy to mention that AXY8 sequences have been experimentally implemented for identifying the NV spin environment~\cite{Hernandez18}, as well as for high-fidelity $^{13}$C nuclear spin control~\cite{Unden19}. In addition, AXY8 sequences have been proposed for positioning nuclear spins in molecules assisted by rf fields~\cite{Wang16}, for quantum information processing in nuclear clusters~\cite{Casanova16} and in trapped-ion systems~\cite{Arrazola18, Dong21}.  

As previously commented in this section, we have introduced the theory of pulsed DD assuming instantaneous pulses (i.e. infinitely energetic pulses). However, any realistic pulse always presents a finite width, which may induce undesired {\it spurious resonances} that in turn can lead to false identification of nuclear species~\cite{Loretz15}. Fortunately, pulsed DD also offers strategies to mitigate these effects. As an example, the randomisation of pulse phases in DD sequences has been proven useful to average out spurious resonances, while at the same it enhances the robustness of the DD method~\cite{Wang19, Wang20}. The natural finite width of the pulses presents other undesired effects, such as the strong reduction of the effective NV-nucleus coupling, that is, the $\frac{f_q A_u^\perp}{4}$ factor in Eq.~(\ref{effectiveNVN}), thus leading to a serious NMR signal-contrast loss. This effect occurs when the pulse duration becomes comparable, or larger, than the nuclear Larmor period as demonstrated in Refs.~\cite{Casanova18, Munuera-Javaloy20}.  Yet, one can resort again to the pulsed DD machinery to overcome this potential problem. Indeed, one can achieve a  maximal NV-nucleus coupling by appropriately modulating the amplitude of the delivered $\pi$ pulses~\cite{Casanova18, Munuera-Javaloy20}. The design of {\it amplitude-modulated $\pi$ pulses} can also incorporate different techniques such as {\it shortcuts to adiabaticity}~\cite{Guery-Odelin19} which results in pulsed schemes that exhibit extra robustness~\cite{Munuera-Javaloy20bis}. In this manner, amplitude-modulated $\pi$ pulses are optimal in situations involving (i) {\it Large static magnetic fields} leading to fast nuclear spin precession and (ii) Scenarios where only a {\it reduced amount of MW power} can be delivered to the sample.  Both situations, (i) and (ii), are interesting from a practical point of view. Large static magnetic fields typically lead to an enhancement of the nuclear decay time, as well as of other effects, such as the chemical shift~\cite{Levitt98}, while the delivery of low MW power pulses allow for the use of pulsed DD techniques over biological samples, as the induced heating is minimal~\cite{Munuera-Javaloy20bis}. 

The type of interactions appearing among NV and nuclei when pulsed DD techniques are employed are of the kind $\sigma_z I^{x}_u$ (for arbitrary $u$),  cf. Eq.~(\ref{effectiveNVN}). Interestingly, this interaction mechanism is not capable of transferring excitations among NV and nuclei, since they do not involve flip-flop terms, in contrast to the one described in Eq.~(\ref{HHHamiltonian}). However, one can combine two pulsed DD sequences, $\sigma_z I^{x}_u$ and $\sigma_z I^{y}_u$, with interspersed $\pi/2$ pulses on the NV to rotate $\sigma_z \rightarrow \sigma_x$ and $\sigma_z \rightarrow \sigma_y$ such that the final interaction is $\propto \sigma_x I^{x}_u +  \sigma_y I^{y}_u \propto (\sigma^+ I^-_u + \sigma^- I^+_u)$. This new interaction term does induce the transfer of NV polarisation to its nuclear environment. In this manner, several pulsed schemes including $\pi$ and $\pi/2$ pulses have been proposed to study hyperpolarisation processes  in diamond. Among them, we can mention the Pulse-Pol sequence~\cite{Schwartz18} that includes {\it robust composite $\pi$ pulses} made of two $\pi/2$ pulses and one central $\pi$ pulse~\cite{Levitt98}. We refer the reader to Ref.~\cite{Schwartz18} for a study of the robustness of Pulse-Pol sequences while, recently, this type of sequences have been also combined with optimal control schemes~\cite{Yang21}. In addition, Pulse-Pol schemes have been recently used for polarising hydrogen spins external to the diamond with an NV ensemble~\cite{Healey21}. See~\cite{Healey21} and references therein for previous developments regarding polarisation of external nuclei.

So far, we have commented on pulsed DD techniques that rely on the perpendicular coupling $A^\perp_j$ among NVs and nuclei. However, one can also exploit the $A^z_j  \sigma_z \vec{I}_j \cdot \hat{z}_j \equiv A^\parallel_j \sigma_z I_j^z$ term in Eq.~(\ref{pulsedH}), which is typically called as {\it parallel coupling}. More specifically, if one combines MW $\pi$ pulses on the NV (leading to $\sigma_z\rightarrow -\sigma_z$) with resonant rf $\pi$ pulses over the $u$th nucleus (with the effect $I^z_u \rightarrow -I^z_u $) the NV-nucleus propagator becomes $e^{-i t A^\parallel_j \sigma_z I_u^z}$. A picture of this radiation scheme can be found in Fig.~\ref{pulsesfig} (c) where it is shown that MW and rf pulses are simultaneously delivered. Note that, the selective rf pulse (that is, a rf pulse resonant with $\omega_u$) is set such that it only rotates the $u$th nucleus, thus the NV coupling with any other nuclei is removed due to the decoupling effect induced by the change $\sigma_z\rightarrow -\sigma_z$ (for more details, see caption of Fig.~\ref{pulsesfig}). This kind of schemes have been used for coupling NVs with fast rotating nuclei~\cite{Aslam17}, as well as with electron-spin radicals encoded in nitroxide labels~\cite{Shi15}. 

\section{Conclusions} In this work we have reviewed distinct dynamical decoupling techniques, both continuous and pulsed, typically used in nanoscale NMR employing NV centers, or other color centers. These techniques have been proven very useful to induce controlled and coherent dynamics among NVs and nuclei and electrons even at room temperature for different tasks,  such as magnetometry or nuclear hyperpolarisation. In this manner, dynamical decoupling methods have a wide applicability and exert a significant impact in different areas of science such as material science, chemistry, and biology.

\acknowledgments
We acknowledge financial support from Spanish Government via PGC2018-095113-B-I00 (MCIU/AEI/FEDER, UE),
Basque Government via IT986-16, as well as from QMiCS (820505) and OpenSuperQ (820363) of the EU Flagship on Quantum Technologies, the EU FET Open Grant Quromorphic (828826) and FET-Open project SuperQuLAN (899354). C.M.-J. acknowledges the predoctoral MICINN grant PRE2019-088519. J. C. acknowledges the Ram\'{o}n y Cajal program (RYC2018-025197-I) and the EUR2020-112117 project of the Spanish MICINN, as well as support from the UPV/EHU through the grant EHUrOPE.

\end{document}